\documentclass[twocolumn]{aastex61}

\usepackage{amsmath}

\newcommand{\vv}{\mbox{\boldmath$v$}}
\newcommand{\vB}{\mbox{\boldmath$B$}}

\shorttitle{Magnetic fields in chondrule-forming shocks}
\shortauthors{Mai et al.}

\begin{document}

\title{Magnetic Fields Recorded by Chondrules Formed in Nebular Shocks}

\correspondingauthor{Chuhong Mai}
\email{chuhong.mai@asu.edu}

\author[0000-0002-9243-5065]{Chuhong Mai}
\affil{School of Earth and Space Exoploration, Arizona State University, Tempe, AZ 85287-1404}

\author[0000-0002-1571-0836]{Steven J. Desch}
\affiliation{School of Earth and Space Exoploration, Arizona State University, Tempe, AZ 85287-1404}

\author{Aaron C. Boley}
\affiliation{Department of Physics and Astronomy, University of British Columbia, 6224 Agricultural Road, Vancouver BC V6T 1Z1 }

\author{Benjamin P. Weiss}
\affiliation{Department of Earth, Atmospheric, and Planetary Sciences, Massachusetts Institute of Technology, 77 Massachusetts Avenue, Cambridge, MA 02139}

\begin{abstract}

Recent laboratory efforts \citep{Fu2014} have constrained the remanent magnetizations of chondrules and the 
magnetic field strengths at which the chondrules were exposed to as they cooled below their Curie points. 
An outstanding question is whether the inferred paleofields represent the background magnetic field of the solar nebula or 
were unique to the chondrule-forming environment. 
We investigate the amplification of the magnetic field above background values 
for two proposed chondrule formation mechanisms, large-scale nebular shocks and planetary bow shocks.
Behind large-scale shocks, the magnetic field parallel to the shock front is amplified by factors $\sim 10-30$,
regardless of the magnetic diffusivity.
Therefore, chondrules melted in these shocks probably recorded
an amplified magnetic field. 
Behind planetary bow shocks, the field amplification is sensitive to the magnetic diffusivity.
We compute the gas properties behind a bow shock around a 3000 km-radius planetary
embryo, with and without atmospheres, using hydrodynamics models.
We calculate the ionization state of the hot, shocked gas, including thermionic emission from dust, and thermal ionization of 
gas-phase potassium atoms, 
and the magnetic diffusivity due to Ohmic dissipation and ambipolar diffusion.
We find that the diffusivity is sufficiently large that magnetic fields have already relaxed to background values  
in the shock downstream where chondrules acquire magnetizations,
and that these locations are sufficiently far from the planetary embryos that chondrules should not have recorded a significant putative dynamo field generated on these bodies.
We conclude that, if melted in planetary bow shocks, chondrules probably recorded the background nebular field.

\end{abstract}

\keywords{meteorites, meteors, meteoroids --- magnetic fields --- planetary systems: protoplanetary disks --- shock waves}

\section{Introduction} \label{sec:intro}

The evolution of protoplanetary disks is strongly affected by magnetic fields.
In weakly magnetized disks---those with background magnetic fields less than about 1 G---the magnetorotational instability (MRI) 
can operate and lead to turbulence and angular momentum transport \citep{Balbus1998, Turner2008}.
In strongly magentized disks---those with background magnetic fields stronger than about 1 G---magnetocentrifugal winds can be 
launched, removing angular momentum from the disk and allowing accretion \citep{Blandford1982, Konigl2000, Wardle2007}. 
Whether one mechanism or the other operates depends on the background magnetic field as well as the magnetic diffusivity.
Regarding the MRI, it is widely appreciated that large zones of protoplanetary disks are probably insufficiently ionized for the 
gas to couple to the magnetic field \citep{Jin1996, Gammie1996, Bai2013, Turner2014}.
The exact coupling of the disk to the magnetic field depends on the diffusivities associated with not just Ohmic dissipation,
but also ambipolar diffusion and the Hall effect (e.g., \citealt{Desch2004}), and therefore is sensitive to the strength and orientation
of the magnetic field.
The rate of MRI-driven accretion is suppressed in weakly magnetized disks and is predicted to be about one order of magnitude lower than observed values. 
Regarding magnetocentrifugal winds, field strengths between about 1 and 30 G are required 
to explain observed stellar accretion rates \citep{Wardle2007}, and the rate at which mass is launched in
the wind depends on the rate of ambipolar diffusion.
Depending on the field strength and other factors, magnetic fields are potentially the dominant driver of protoplanetary disk 
evolution.
Yet very little data exist on the magnetic field strength in extant protoplanetary disks or the solar nebula.
While observations of the polarization of millimeter emission are beginning to reveal the geometry of magnetic fields in 
protoplanetary disks \citep{Stephens2014, Li2016}, the strength of the magnetic field is only poorly constrained
to be $> 0.01 \, {\rm G}$ \citep{Li2016, Crutcher2012}.

Only recently have new laboratory analyses of meteorites begun to open a window into the magnetic field of the solar nebula. 
Highly sensitive measurements \citep{Fu2014, Fu2015} have revealed the remanent magnetization of chondrules, igneous inclusions
in meteorites that formed in the solar nebula.
Chondrules are millimeter- to submillimeter-sized igneous droplets found in abundance in chondritic meteorites. 
They are primarily composed of ferromagnesian silicates (olivines and pyroxenes), but often contain other minerals such as Fe-sulfide
(troilite) or metallic FeNi (kamacite). 
Textural and chemical evidence strongly indicates that chondrules were ``flash-heated" within minutes from temperatures below 650 K 
to temperatures above the liquidus, becoming free-floating molten objects in the presence of solar nebula gas; over the course of hours 
they cooled and crystallized \citep{Lofgren1990, Rubin1999, Connolly2004, Lauretta2006, Desch2012}.
The heating events that melted precursors and formed most chondrules took place in the protoplanetary disk, about 1.5 to 4.0 Myr 
after the formation of calcium-rich, aluminum-rich inclusions (CAIs), the first solids to form in the solar nebula 
\citep{Kurahashi2008}. 
As chondrules crystallized and cooled from the Curie points of their constituent ferromagnetic minerals within them (e.g., the Curie point of
kamacite is 1038 K) to ambient space temperatures, the strength of the magnetic field in the chondrule-forming region would be continuously recorded as thermoremanent magnetization 
by those minerals.
About 10\% of chondrules found in the Semarkona LL3.0 chondrites have "dusty olivines", olivine phenocrysts containing kamacite 
inclusions capable of retaining magnetizations \citep{Fu2014, Fu2015}.
Magnetizations of different directions have been recorded in the Semarkona chondrules, indicating that chondrules were exposed to
magnetic fields before accretion into the Semarkona parent body \citep{Fu2014, Fu2015}.
These studies show that the strength of the magnetic field chondrules were exposed to on average was $\approx 0.54 \pm 0.21$ G.
In contrast, the magnetizations of chondrules in the CR chondrite LAP02342 may indicate a weaker magnetic field $< 0.15$ G
\citep{Fu2015}.
These measurements improved on earlier attempts that did not test for random orientations of the magnetizations that would
indicate a pre-accretionary paleofield \citep{Fu2014}.

Although the magnetic field strengths {\it in the chondrule-forming region} are beginning to be constrained by meteoritic analyses,
it is not yet clear how these relate to the background magnetic field of the solar nebula.
The relationship between the background field and the paleofield in the chondrule-forming region depends on the exact chondrule
formation mechanism. 
Several theories of chondrule formation have been proposed.
A successful model of chondrule formation must be tested against experimental and other constraints regarding the timing, frequency,
and physical conditions of chondrule formation, as well as the thermal histories experienced by chondrules (see \citealt{Desch2012} for
a review). 
Some of the well-developed models include: ejection of molten chondrules by impact between asteroids 
(e.g., \citealt{Urey1953, Asphaug2011, Johnson2015});
melting of chondrules in flares near the early Sun, as in the ``X-wind" model 
\citep{Shu1996, Shu1997, Shu2001}; 
melting by lightning in the solar nebula \citep{Morfill1993, Pilipp1998, Desch2000}; 
melting of chondrules by large-scale shocks in the solar nebula, such as those produced by gravitational instabilities 
(e.g., \citealt{Wood1984, Wood1996, Desch2002, Boley2008, Morris2010});
or small-scale shocks in the solar nebula, such as bow shocks produced by planetesimals 
(e.g., \citealt{Hood1998, Ciesla2004, Hood2009}) or planetary embryos 
\citep{Morris2012, Boley2013, Mann2016} on eccentric orbits. 
The magnetic field in the chondrule-forming region may reflect the background value of the solar nebula, or not, depending on how
chondrules actually formed, and depending on the geometry of the magnetic field and the rate of magnetic diffusivity in the 
chondrule-forming region. 

For example, in the impact model, chondrules rapidly leave the vicinity of a parent body and can be expected to record the background
field of the nebula. 
Chondrules formed according to the ``X wind" model would be melted very near ($\sim 0.1$ AU) from the early Sun, in a unique environment
near the magnetospheric radius, where the magnetic field geometry drives a magnetocentrifugal outflow. 
Though \cite{Shu1996} predicted fields of 0.4 - 8 G for the location where X-wind produced chondrules cooled,
chondrules melted here could
be exposed to fields of 10 G or more \citep{Bai2013}, and these paleofields would not represent the overall background field of
the protoplanetary disk. 
If chondrules were melted by nebular lightning, it is unclear what paleofields they would record.  
While fulgurites (soils melted by lightning strikes) record magnetic fields associated with the lightning strike itself ($>$ 1 T), through
the process known as lightning-induced remanent magnetization (e.g. \citealt{Sakai1998, Sakai2002, Salminen2013}), it is not at all clear that the magnetic fields associated 
with solar nebula lightning currents can be maintained for the hours it took chondrules to cool.
As such, chondrules melted by lightning probably would record the background magnetic field, but that is uncertain.
Finally, in the case of chondrule formation by solar nebula shocks, chondrule precursors are melted as they pass through the 
shock front and they cool and crystallize in the post-shock region.
As we show below in \S~\ref{sec:lsns}, 
in the post-shock region of a large-scale, 1-D shock, the magnetic field is amplified above the background magnetic 
field of the nebula by a factor $\chi$ that varies from $\chi \sim 1$ for shocks propagating parallel to the magnetic field, to 
$\chi \sim {\cal M}^2 \sim 30$ (where ${\cal M}$ is the Mach number) for shocks propagating across field lines.
These results are independent of the magnetic diffusivity.
In contrast, in a small-scale, effectively 2-D shock, magnetic diffusivity can play an important role.
If magnetic diffusivity is low, the post-shock  magnetic field can remain amplified by a factor $\chi$ as in a 1-D shock;
but if magnetic diffusivity is high, the magnetic flux amplified behind the shock can diffuse laterally, returning the magnetic 
field in the post-shock region to the background value.

In this paper, we examine in greater detail the post-shock evolution of the magnetic field behind planetary embryo bow shocks.
The paleofields recorded by chondrules melted by mechanisms other than shocks are simple to interpret: chondrules melted by 
impacts or lightning presumably record the background field of the nebula, and chondrules melted by the X wind mechanism
presumably record the magnetic field close to the early sun. 
The paleofields recorded by chondrules melted by large-scale 1-D shocks are analytically simple and is considered in \S~\ref{sec:lsns}.
The last example, of chondrules melted in small-scale planetary embryo bow shocks, is more complicated and is the focus of
the present paper.
In \S~\ref{subsec:3dmodel} we discuss the temperature, pressure and density of gas behind the shock of an atmosphere-free planetary embryo, for which we draw on the results 
of modeling using the hydrodynamics code Bozxy Hydro \citep{Boley2013}.
Because the magnetic diffusivity is important in this case, we calculate in \S~\ref{subsec:ionizn} the ionization of gas in the post-shock 
region using the formalism in \cite{Desch2015}, including not only thermal ionization of gas-phase potassium, but also thermionic 
emission of electrons and ions from hot dust grains.
In \S~\ref{subsec:diffu} we use these computed ionization fractions to calculate the magnetic diffusivities associated with 
Ohmic dissipation and ambipolar diffusion.
We compare the calculated magnetic diffusivities with the simulated trajectories of chondrules through the bow shock in \S~\ref{subsec:results} and show that the majority of chondrules could not record an amplified paleofield upon cooling down from the Curie point.
We also show that the chondrules cool sufficiently far away from the embryo such that if its metallic core was generating a dynamo, this would impart only a relatively weak magnetization on the chondrules relative to the nebular field.
In \S~\ref{subsec:atmos}, we perform similar calculations to a protoplanet with a thick atmosphere. 
We show that most chondrules cool down far downstream and avoid the region where magnetic field can stay amplified
and also avoid any embryo dynamo fields.
We conclude in \S~\ref{sec:condis} that if chondrules are melted by planetary embryo bow shocks, then they record the background magnetic field of 
the solar nebula. 
We also discuss the implications for the protoplanetary disk evolution. 
\\

\section{Large-Scale (1-D) Nebula Shocks} 
\label{sec:lsns}

Before considering the more complicated case of 2-D planetary embryo bow shock, we first consider the evolution of the magnetic
field behind a large-scale 1-D shock. 
Large-scale shock waves are consistent with most of the experimental constraints on chondrule formation and are a leading model
for chondrule formation \citep{Desch2012}.
Potentially they can be generated by accretion shocks \citep{Wood1984, Ruzmaikina1994}, 
infalling clumps \citep{Tanaka1998}, or tidal effects from passing stars \citep{Larson2002}.
Large-scale shocks at the disk midplane, which are most consistent with the constraints on chondrule formation, 
can be generated readily by global gravitational instabilities \citep{Wood1996, Boss2005}.
Such shocks have lateral extent $\sim 1$ AU, comparable to the size of the disk, much greater than the effective thickness 
of the shock, $\sim 10^5$ km \citep{Desch2002}. They are effectively 1-D.
Due to the simplified geometry, the evolution of the magnetic field can be calculated analytically. 

In the flux-freezing limit, the behavior of the magnetic field is especially straightforward to calculate. 
The gas motion is described by the mass continuity equation: 
\begin{equation} 
\frac{\partial \rho}{\partial t} + \nabla \cdot \left( \rho \vv \right) = 0,
\end{equation} 
or, using the advective derivative,  
\begin{equation} 
\frac{d \rho}{dt} = -\rho \, \left( \nabla \cdot \vv \right).
\end{equation} 
In addition to this equation is the magnetic field evolution equation in the flux freezing approximation: 
\begin{equation}
\frac{\partial \vB}{\partial t} - \nabla \times \left( \vv \times \vB \right) = 0.
\end{equation}
Using $\nabla \cdot \vB = 0$ and rewriting using the advective derivative yields
\begin{equation} 
\frac{d \vB}{dt} = \vB \cdot \nabla \vv - \vB \, \left( \nabla \cdot \vv \right). 
\end{equation} 
Combining, one finds
\begin{equation}
\frac{d}{dt} \left( \frac{ \vB }{ \rho } \right) = \frac{1}{\rho} \, \vB \cdot \nabla \, \vv
\end{equation}
For the case that the magnetic field is perpendicular to the velocity gradients, the right-hand side vanishes, 
and one finds that the magnetic field does not change direction, and its strength is proportional to the density, $\rho$. 
For the case where the gradients and the flow direction are along the magnetic field, it is straightforward to
show that the magnetic field is constant and uniform. 

These equations are even simpler in the context of a steady-state, 1-D shock. 
We consider a 1-D shock in which $\vv$ is along the $x$ direction, $\vB$ is along the $z$ direction, and all 
variations are along the $x$ direction.
In that case, assuming a steady-state solution, we must have 
\begin{equation}
\frac{\partial}{\partial x} \left( v B \right) = 0, 
\end{equation}
leading to the ``jump condition" across the shock front:
\begin{equation}
v_{\rm pre} \, B_{\rm pre} = v_{\rm post} B_{\rm post}.
\end{equation}
Note that velocities are defined relative to the frame of the shock. 
The equation of mass conservation yields a very similar formula relating the density before and after the shock:
\begin{equation}
\rho_{\rm pre} \, v_{\rm pre} = \rho_{\rm post} v_{\rm post},
\end{equation}
so we can write
\begin{equation}
\frac{ B_{\rm post} }{ B_{\rm pre} } = \frac{ v_{\rm pre} }{ v_{\rm post} } = 
 \frac{ \rho_{\rm post} }{ \rho_{\rm pre} } \equiv \chi.
\end{equation}
For an adiabatic gas with adiabatic index $\gamma$, 
\begin{equation}
\chi = \frac{ (\gamma+1) \, {\cal M}^2 }{ (\gamma-1) {\cal M}^2 + 2},
\end{equation}
where 
\begin{equation}
{\cal M}^2 = \frac{ \bar{m} V_{\rm s}^2 }{ \gamma k T_{\rm pre} }
\end{equation}
is the Mach number squared, $V_{\rm s}$ being the shock speed (e.g., \citealt{Shore1992}).
In the strong-shock limit (${\cal M}^2 \gg 1$), $\chi \approx (\gamma+1) / (\gamma-1) \approx 6$ typically.
But if the gas can radiate, then $\chi$ can reach higher values: in an isothermal shock the 
compression factor is $\chi \approx 2\gamma / [(\gamma+1) \, {\cal M}^2]$ and is technically unlimited.
The calculations of \cite{Desch2002} find $\chi \sim 11$ for the shocks most likely to produce
chondules, at the post-shock location where chondrules reach the Curie point. 
We consider $\chi \approx 10-30$ to be represent the typical compression of the magnetic field. 

Because solar nebula gas is only partially ionized, the assumption of flux freezing is not always justified, and we 
must include non-ideal magnetohydrodynamic effects in the magnetic field evolution equation, as follows:
\begin{equation}
\frac{\partial \vB}{\partial t} -\nabla \times \left( \vv \times \vB \right) = 
-\nabla \times \left( {\cal D} \, \nabla \times \vB \right),
\end{equation}
where ${\cal D} = (c^2 / 4\pi) \eta_{\perp}$ is the coefficient of magnetic diffusion (and $\eta_{\perp}$ the resistivity)
associated with currents perpendicular to the magnetic field \citep{Parks1991}.
This diffusion coefficient includes the effects of Ohmic dissipation and ambipolar diffusion, and depends in a complicated
way on the ionization fraction, the density and temperature, and the local magnetic field strength.
The Hall diffusion term can be neglected here as we are mainly considering a poloidal magnetic field in the background.
We revisit the calculation of ${\cal D}$ in \S~\ref{subsec:diffu}.
For now we assume that ${\cal D}$ is uniform in the pre-shock region. 

If we assume steady state and the same 1-D geometry as above, we must have 
\begin{equation}
\frac{\partial}{\partial x} \, \left( v B - {\cal D} \, \frac{\partial B}{\partial x} \right) = 0.
\end{equation}
It is straightforward to show that the solution to this equation is 
\begin{eqnarray}
B_{\rm pre}(x) & = & B_0 + B_0 \, \left( \chi - 1 \right) \, e^{+x / L} \\
B_{\rm post}   & = & \chi \, B_0,
\end{eqnarray}
where $B_0$ is the background magnetic field strength in the solar nebula, 
$L = {\cal D} / V_{\rm s}$ is a characteristic diffusion lengthscale,
and the pre-shock region is defined by $x < 0$.
The magnetic field is compressed by the shock by a factor of $\chi$. 
Because magnetic flux cannot diffuse in the lateral direction, it remains 
compressed in the post-shock region.
Magnetic flux does diffuse into the pre-shock region, but the diffusion of
this flux is exactly cancelled by the advection of magnetic flux into the 
shock by the supersonic gas. 
The surprising result is that chondrules melted by large-scale shocks will
record the amplified magnetic field, not the background magnetic field,
regardless of the magnetic diffusivity (unless the shock velocities are 
exactly along the magnetic field lines).

\section{Planetary Embryo Bow Shocks} \label{sec:pbs}

Large-scale nebular shocks are a mechanism consistent with almost all experimental and other constraints on chondrule formation.
The bow shocks around planetary embryos are also largely consistent with the constraints
\citep{Morris2012, Boley2013, Mann2016}. 
The basic idea behind the model is that planetary embryos $> 2500 \, {\rm km}$ in radius apparently existed in the solar nebula
(Mars had largely formed by 1.8 Myr: \citealt{Dauphas2011}). 
If any these embryos were scattered or perturbed onto eccentric orbits ($e > 0.25$ or so), then they would move supersonically,
at speeds $\sim 8 \, {\rm km} \, {\rm s}^{-1}$, with respect to the gas \citep{Morris2012}. 
They would therefore drive a bow shock in front of them, and chondrule precursors passing through this shock front would
be flash-heated and melted in ways consistent with constraints.
Chondrule precursors first are heated as the shock front approaches, by infrared radiation emitted from the already-heated
chondrules behind the shock front.
Then the chondrules receive an intense pulse of frictional heating as they pass through the shock front and move supersonically 
with respect to the gas.
After the aerodynamic stopping time (about 1 minute), the chondrules no longer have any relative velocity with respect to the gas. But they are still heated
by thermal exchange with the hot, shocked gas, as well as by infrared radiation emitted by other chondrules. 
Eventually the planetary body moves on, the system cools, and the melted precursors cool and crystallize, forming chondrules.

The planetary embryo bow shock model is inspired by and resembles earlier models of chondrule formation in planetesimal bow shocks
\citep{Hood1998, Ciesla2004, Hood2005, Hood2009}.
The main difference is that planetary embryos are $\sim 2500$ km in radius, as opposed to tens to hundreds of km in radius.  
But the larger scale makes qualitative differences. For one thing,
only larger planetary bodies could be perturbed into highly eccentric orbits as needed for bow shock formation \citep{Hood2012}.
For another, the planetary embryo bow shock model does not suffer from 
deficiencies of planetesimal bow shock models.
The most severe of these is that any chondrules created by planetesimal bow shocks will be immediately accreted onto the body.
For planetary embryos, the larger scale yields a larger stand-off distance between the shock front and the planetary surface,
and only for embryos does the stand-off distance exceed the aerodynamic stopping distance. 
It is also the case that, as first pointed out by \cite{Ciesla2004}, peak temperatures in planetesimal bow shocks ($<$ 1000 km) are too low and cooling rates are too rapid to match the thermal histories of chondrules. In fact, the cooling rates in these small scale shocks are about an order of magnitude faster than the already
high cooling rates in planetary embryo bow shock models.

In this section we calculate the rates of magnetic diffusion behind a planetary embryo bow shock.
We construct a model for the gas density and temperature using the output of bow shock models similar to those in \cite{Boley2013}. 
We then calculate the ionization state using the prescriptions of \cite{Desch2015}.
We convert these ionizations into rates of magnetic diffusion, and then we compare the implied diffusion timescales to 
the dynamical timescales throughout the post-shock region.
We show that in the regions in which chondrules will cool below their Curie points, the rates of magnetic diffusion are
sufficiently high that the magnetic field strength should relax to the pre-shock, background value.

\begin{figure*}[ht!]
\centering
\includegraphics[scale=0.55]{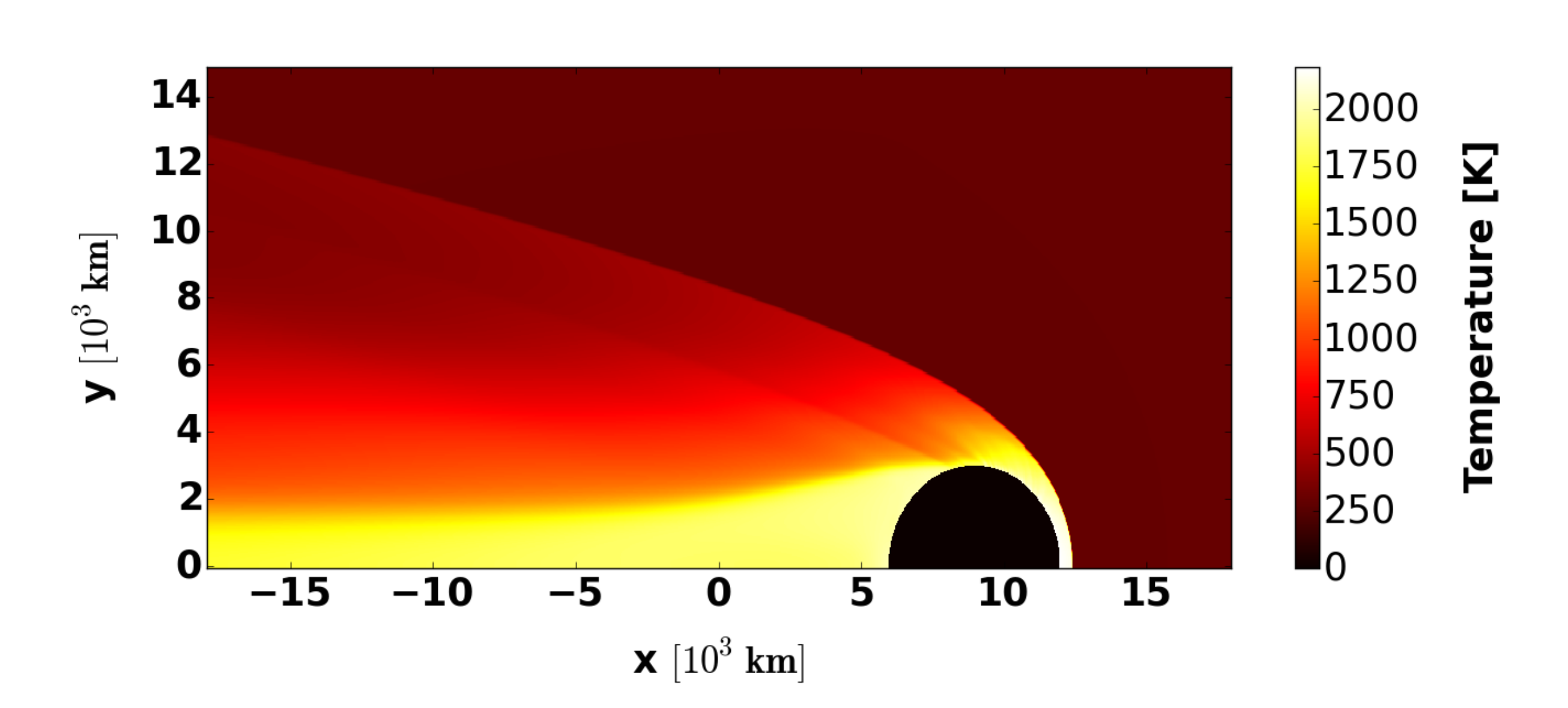}
\includegraphics[scale=0.55]{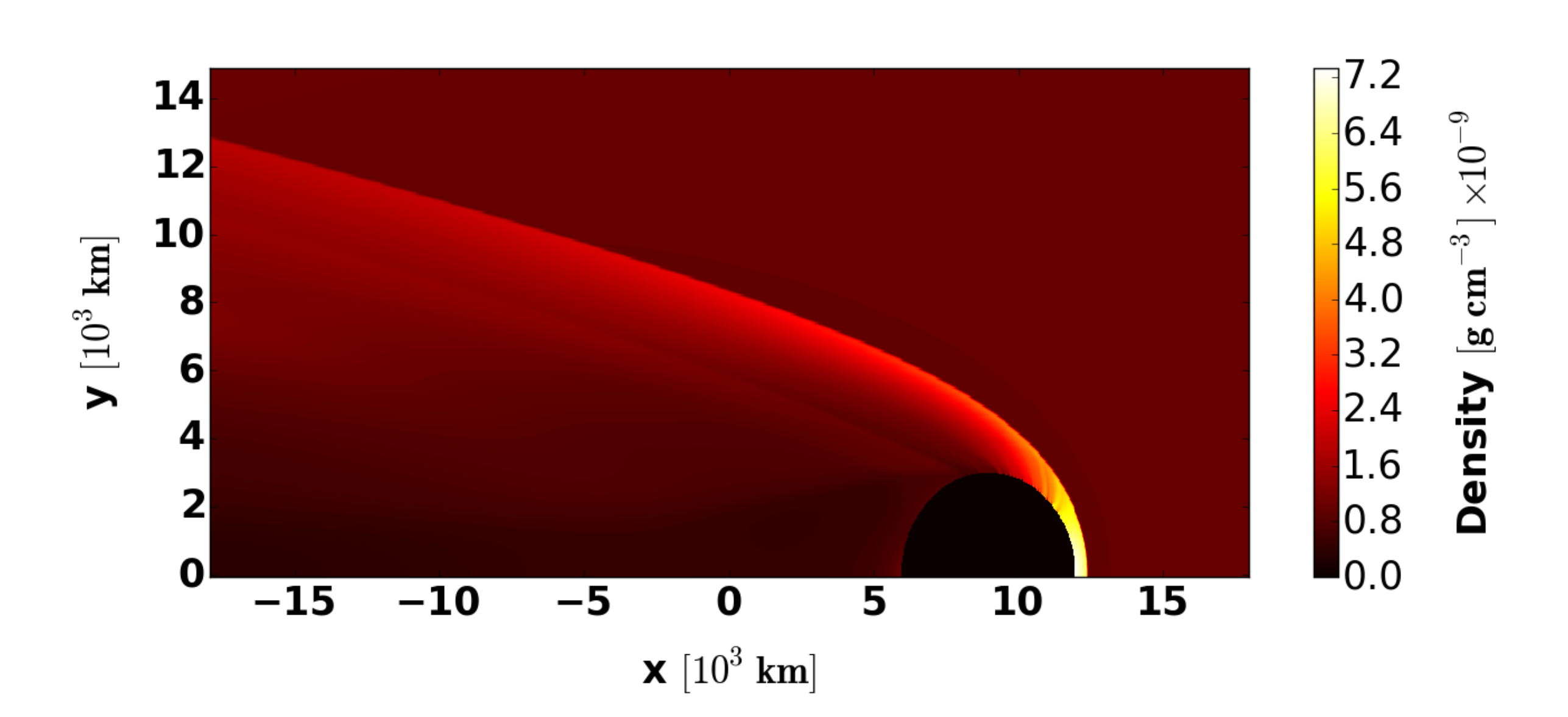}
\caption{Representative gas temperature (top) and density (bottom) in the planetary bow shock region. 
The embryo is drawn as black. \label{fig:temp}}
\end{figure*}

\subsection{3-D Bow Shock Model} 
\label{subsec:3dmodel}

We first establish the background conditions using output obtained in the 2-D/3-D planetary embryo bow shock models (i.e., we use a 2-D grid that captures 3-D effects using cylindrical cells with corresponding updates the hydrodynamics).
These simulations use output from the 3-D, second-order accurate radiation hydrodynamics code called Boxzy Hydro \citep{Boley2013}.
The Cartesian grid includes 960 cells along the $x$ axis and 400 cells along the other directions, with a resolution of 37.4 km.
The planetary embryo, with radius 3000 km and mass $3.4\times 10^{26}\ {\rm g}$, is held stationary in these simulations, while gas is allowed to flow along the $x$ axis, 
entering the computational domain with a uniform density $10^{-9} \, {\rm g} \, {\rm cm}^{-3}$, temperature 300 K, and velocity 
$V_{\rm s} = 7 \, {\rm km} \, {\rm s}^{-1}$. 
The gas is composed of 73 wt\% ${\rm H}_{2}$ and 25 wt\% He, with heavier elements comprising 2\% of the mass, yielding a mean
molecular weight 2.33 amu. 
This code adopts an equation of state that includes the rotational and vibrational states of H$_2$, as well as H$_2$ dissociation and recombination.
In the particular run employed here, the ability to radiate is turned off, so that the gas behaves adiabatically. 
This is an optimistic scenario because gas tend to cool faster if radiating. For the particular science explored here, this assumption simplifies the simulation while leaving the nature of problem unchanged.
The code uses particle-in-cell methods for particle integration to keep track of dynamical feedbacks from solids on the gas. 
To simulate the trajectories of chondrule precursors passing through the bow shock, we inject $10^6$ super-particles 
with random coordinates into the simulation domain, together with the wind flowing along the $x$ axis.
The solid-to-gas ratio is kept to be $\sim 0.004$, which is consistent with the average chondrule mass fraction ($\sim 0.04$) and chondrule concentration factor ($\sim 10$) for chondrule-forming regions in previous studies \citep{Morris2012, Boley2013}.
40 tracer particles with evenly increasing impact parameters up to 6000 km are selected from these super-particles.
Their positions and corresponding physical properties (temperature, density, etc.) are regularly recorded during the simulation.
The code is evolved for $\sim 10^6$ s until a quasi-steady state is achieved.
Major outputs include pressure, temperature, gas density and velocity distribution. 
Fig. \ref{fig:temp} shows the representative temperature and gas density around the modeled planetary embryo.

\subsection{Ionization States}
\label{subsec:ionizn}

The degree to which gas couples dynamically to magnetic fields is determined by the magnetic diffusion rates, which in turn
depend on the degree of ionization, among other factors. 
The rate of ionization of gas in disk midplanes is remarkably low, as energetic particles (Galactic cosmic rays, solar energetic
particles, and X rays and UV photons from the star) are shielded by the disk gas. 
Essentially the only source of ionization in cold gas is from radioactive decay, e.g., of ${}^{40}{\rm K}$, ${}^{238}{\rm U}$, ${}^{235}{\rm U}$ and ${}^{232}{\rm Th}$, at a rate $\zeta \sim 10^{-22} \, {\rm s}^{-1}$ \citep{Umebayashi2009} per H$_2$ molecule, and of short-lived radionuclide ${}^{26}{\rm Al}$, 
$\zeta \sim 10^{-18} \, {\rm s}^{-1}$  \citep{Umebayashi1981}.
Ionization rate increases considerably with temperature as the gas heats above about 700 K, however, due to thermionic emission
from hot solids, and at even higher temperatures from collisional ionization of gas-phase alkali atoms \citep{Desch2015}. 
Traditionally, only the latter effect has been considered in disks, assuming that the rates of collisional ionization of alkalis, 
\[
{\rm K}^{0} + {\rm H}_{2} \rightarrow {\rm K}^{+} + {\rm e}^{-} + {\rm H}_2,
\]
is in detailed balance with the rate of gas-phase recombinations, 
\[
{\rm K}^{+} + {\rm e}^{-} + {\rm H}_2 \rightarrow {\rm K}^{0} + {\rm H}_{2} 
\]
(where K is the alkali element that contributes most strongly to ionization).
If these were in balance, then one could apply the equilibrium Saha equation to calculate the ionization fraction as a function
of temperature:
\begin{equation} 
\label{eq:saha}
 \frac{n_{{\rm K}^{+}}}{n_{{\rm K}^0}} = \frac{g_+}{g_0} \, \frac{2}{n_{\rm e}} \, 
 \left( \frac{2\pi m_{\rm e} \, k T}{h^2} \right)^{3/2} \, \exp \left( -\frac{ {\rm IP} }{ k T } \right),
\end{equation}
where $n_{K^+}$, $n_{K^0}$ and $n_e$ are densities of potassium ions and atoms, and electrons, $g_+ = 1$ and $g_0 = 2$ are 
statistical weights of ${\rm K}^+$ and ${\rm K}^0$, $m_{\rm e}$ is the electron mass, $k$ is Boltzmann's constant, 
$h$ is Planck's constant, and ${\rm IP} = 4.34 \, {\rm eV}$ is the first ionization potential energy of potassium. 
In fact, as demonstrated by \cite{Desch2015}, the recombination of alkali ions is primarily on dust surfaces, as ions and 
electrons that adsorb onto the same grain can quantum tunnel over the grain surface until they recombine (and most likely leave as 
a neutral atom).
Sticking of electrons or ions on grain surfaces is in fact in detailed balance with emission of electrons or ions from
grain surfaces by the process of thermionic emission, at a rate given by Richardson's law: 
\begin{equation}
 j(T) = \lambda_{\rm R} \frac{4\pi m_{\rm e} (k T)^2 }{ h^3 } \, \exp \left( -\frac{W}{k T} \right),
\end{equation}
where $j(T)$ is temperature-dependent rate of electron emission per grain surface area, $\lambda_{\rm R}$ is a material-dependent
dimensionless constant, ${W}$ is the work function of the solid, and other constants have their usual meanings. 
Because the work functions of astrophysical solids typically are $\approx 5 \, {\rm eV}$, the ionization due to thermionic
emission approximates that of thermal ionization, especially above about 1500 K, but differs significantly at lower temperatures
\citep{Desch2015}.
The exponential factor present in both the Saha equation and Richarson's law highlights the sensitivity of the ionization
to temperature.

\begin{figure*}[ht!]
\centering
\includegraphics[scale=0.55]{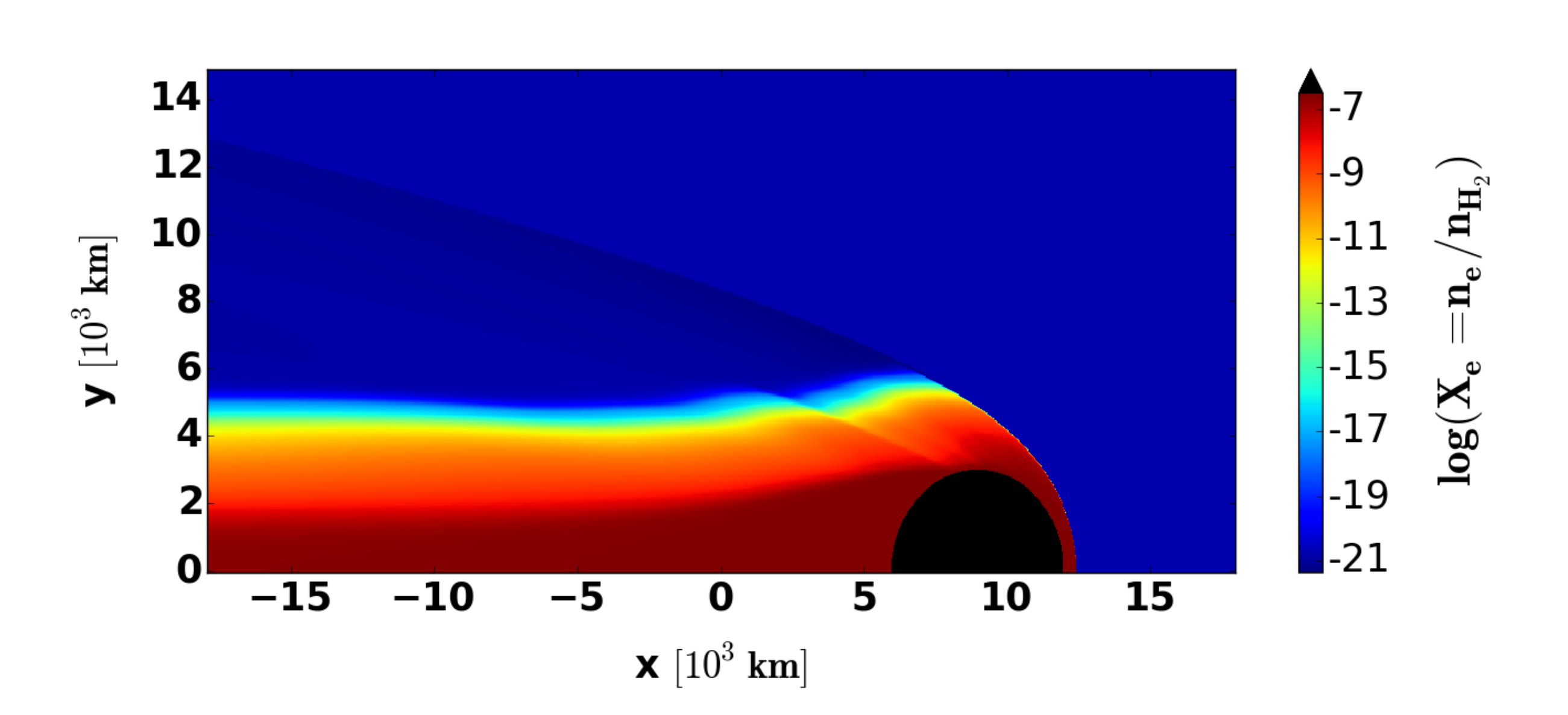}
\includegraphics[scale=0.55]{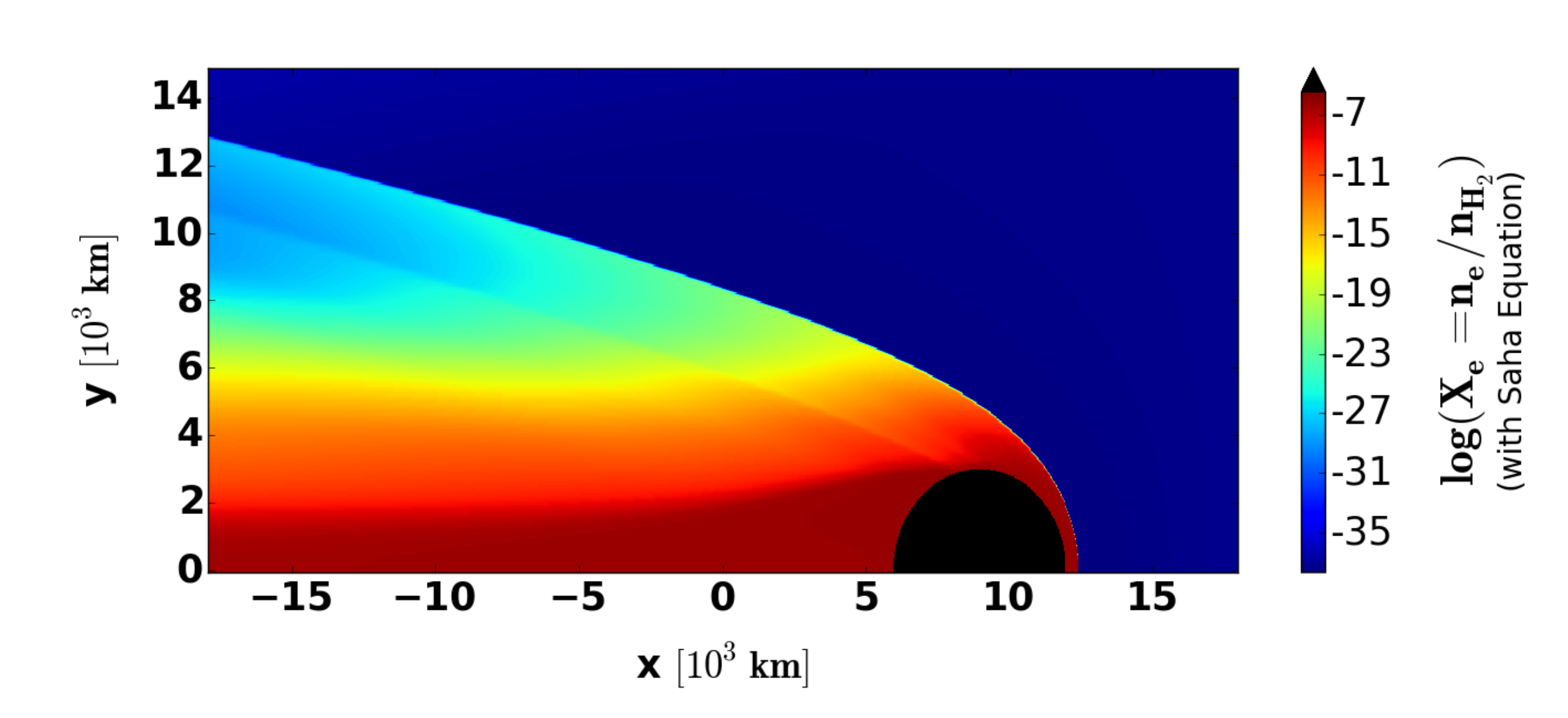}
\caption{Electron fraction in the post-shock gas behind the planetary embryo bow shock. The post-shock region has a relatively 
high ionization fraction, especially directly behind the planetary body (drawn as a black circle).
The upper panel considers thermionic emission from hot dust grains while the lower panel shows the results using the Saha equation, as a comparison. See text for discussion.
\label{fig:ioniz}}
\end{figure*}

We take the temperature from the output of the planetary bow shock models; as seen in Fig.~\ref{fig:temp}, the temperature ranges from 
300 K to over 2000 K in the bow shock region. 
We adopt the chemical network built by \cite{Desch2015}, which includes thermal ionization, gas-phase recombination, adsorption 
following collisions onto dust grains, and thermionic emission of electrons and ions (${\rm K}^{+}$) and neutral ${\rm K}^{0}$ atoms
from grain surfaces. 
This formalism returns the steady-state densities of free electrons (${\rm e}^{-}$), neutral and ionized gas-phase atoms 
(${\rm Mg}^{+}$), neutral and ionized gas-phase alkali atoms (${\rm K}^{0}$, ${\rm K}^{+}$, and grain-adsorbed alkali atoms
(${\rm K}^{0*}$), as well as the densities of neutral and charged grains (${\rm g}^{+}$, ${\rm g}^{0}$, ${\rm g}^{-}$), as a
function of density ($\rho$), temperature (T), abundances and grain mass fraction, and background non-thermal ionization rate
$\zeta$. 
The steady-state assumption is justified at the higher temperatures of interest: for $T > 1500 \, {\rm K}$, ionization equilibrium
is achieved in less than 2-3 minutes, and in less than an hour for $T > 1200 \, {\rm K}$ (\citealt{Desch2015}, Figure 5).
Rather than run the code at every location, we ran it separately to generate a lookup table that returns the above densities 
as a function of temperature and hydrogen number density ($n_{\rm H} = \rho / (2.8 \, m_{\rm H})$, to account for He).
We assume a solids-to-gas mass ratio of 0.004, a grain radius $1 \, \mu{\rm m}$ and internal density $3 \, {\rm g} \, {\rm cm}^{-3}$,
and a particle work function $W = 5.0 \, {\rm eV}$ appropriate for ferromagnesian silicates.
We find that for the conditions typical of the high-temperature regions behind the bow shock, the dominant ionization mechanism is
emission from dust surfaces, and the dominant recombination mechanism is adsorption onto dust surfaces, so the ionization state is 
relatively insensitive to the exact details of particle size or solids-to-gas mass ratio. 
Fig.~\ref{fig:ioniz} upper panel shows the electron fraction $n_{\rm e} / n_{\rm H}$ as a function of location behind the shock. 
Densities of other charged particles follow similar spatial distributions.
As a comparison, we also used the Saha equation (Eqn. \ref{eq:saha}) to find the electron fraction (Fig.~\ref{fig:ioniz} lower panel).
The results deviate from the lookup table at low temperatures, as the Saha equation predicts much smaller ionization fraction in those regions. 
At temperatures around 1500 $\sim$ 2000 K, the ionization state 
around the planetary bow shock is adequately described by the Saha equation.
\\

\subsection{Magnetic Diffusion}
\label{subsec:diffu}

For the case that magnetic fields are perfectly coupled to the fluid (the flux-freezing approximation), 
The main determinant of how well magnetic fields couple to the gas is the degree of ionization, but other factors matter as well,
especially magnetic field strength and geometry, and density and temperature. 
We consider Ohmic dissipation and ambipolar diffusion as diffusion mechanisms: the Hall effect can sometimes dominate, but under
almost all conditions in protoplanetary disks, either Ohmic dissipation or ambipolar diffusion dominates \citep{Desch2015}. 
The diffusion rate depends on the magnetic field geometry: 
if the electric currents are parallel to the magnetic field, the diffusion coefficient ${\cal D}_{\parallel}$ is proportional 
to the parallel resistivity, $\eta_{\parallel}$, which is due entirely to Ohmic dissipation: $\eta_{\parallel} = \eta_{\rm OD}$.
If the electric currents are perpendicular to the magnetic field, the diffusion coefficient ${\cal D}_{\perp}$ is proportional to 
the perpendicular resistivity, $\eta_{\perp}$, which is the sum of contributions from Ohmic dissipation and ambipolar diffusion:
$\eta_{\perp} = \eta_{\rm OD} + \eta_{\rm AD}$ (e.g., \citealt{Desch2001}). 
We do not calculate the magnetic field geometry, so we cannot quantify the relative contributions of Ohmic dissipation and 
ambipolar diffusion to the uncoupling of the magnetic field from the gas; therefore we consider the rate of magnetic diffusion to
be bracketed by ${\cal D}_{\parallel}$ and ${\cal D}_{\perp}$. 

The magnetic diffusion coefficients are related to the resistivities as ${\cal D}_{\parallel} = (c^2 / 4\pi) \, \eta_{\parallel}$
and ${\cal D}_{\perp} = (c^2 / 4\pi) \eta_{\perp}$, where the resistivities are related to the conductivities, $\sigma$, as 
\begin{equation}
\eta_{\parallel} = \eta_{\rm OD} = \frac{1}{\sigma_{\parallel}} 
\end{equation} 
and
\begin{equation}
\eta_{\perp} = \eta_{\rm OD} + \eta_{\rm AD} = \frac{\sigma_{\perp}}{ \sigma_{\perp}^2 + \sigma_{\rm H}^2 },
\end{equation} 
where 
\begin{eqnarray} 
\sigma_{\parallel} & = &  \sum_{s} \sigma_{s} \\ 
\sigma_{\perp}     & = &  \sum_{s} \frac{\sigma_s}{1 + (\omega_s \tau_{sn})^2} \\ 
\sigma_{\rm H}     & = & -\sum_{s} \frac{\sigma_s \, \omega_s \tau_{sn}}{1 + (\omega_s \tau_{sn})^2},
\end{eqnarray}
where $\sigma_{\parallel}$ is the direct conductivity, $\sigma_{\perp}$ the Pedersen conductivity, and 
$\sigma_{\rm H}$ the Hall conductivity.
Here $\sigma_{s} = n_s q_s^2 \tau_{sn} / m_s$ is the conductivity of each species $s$, 
where $n_s$, $q_s$ and $m_s$ are the number density, charge and mass of each charged species, 
$\omega_{s} = q_s B / m_s c$ is the gyrofrequency of species $s$ around magnetic field lines, and 
$\tau_{sn}$ is the momentum exchange timescale for charged species $s$ in a sea of neutral particles,
formulas of which are taken from \cite{Pinto2008}.
The conductivities sum over the charged species ${\rm e}^{-}$, ${\rm Mg}^{+}$ and ${\rm K}^{+}$.

\begin{figure*}[ht!]
\centering
\includegraphics[scale=0.55]{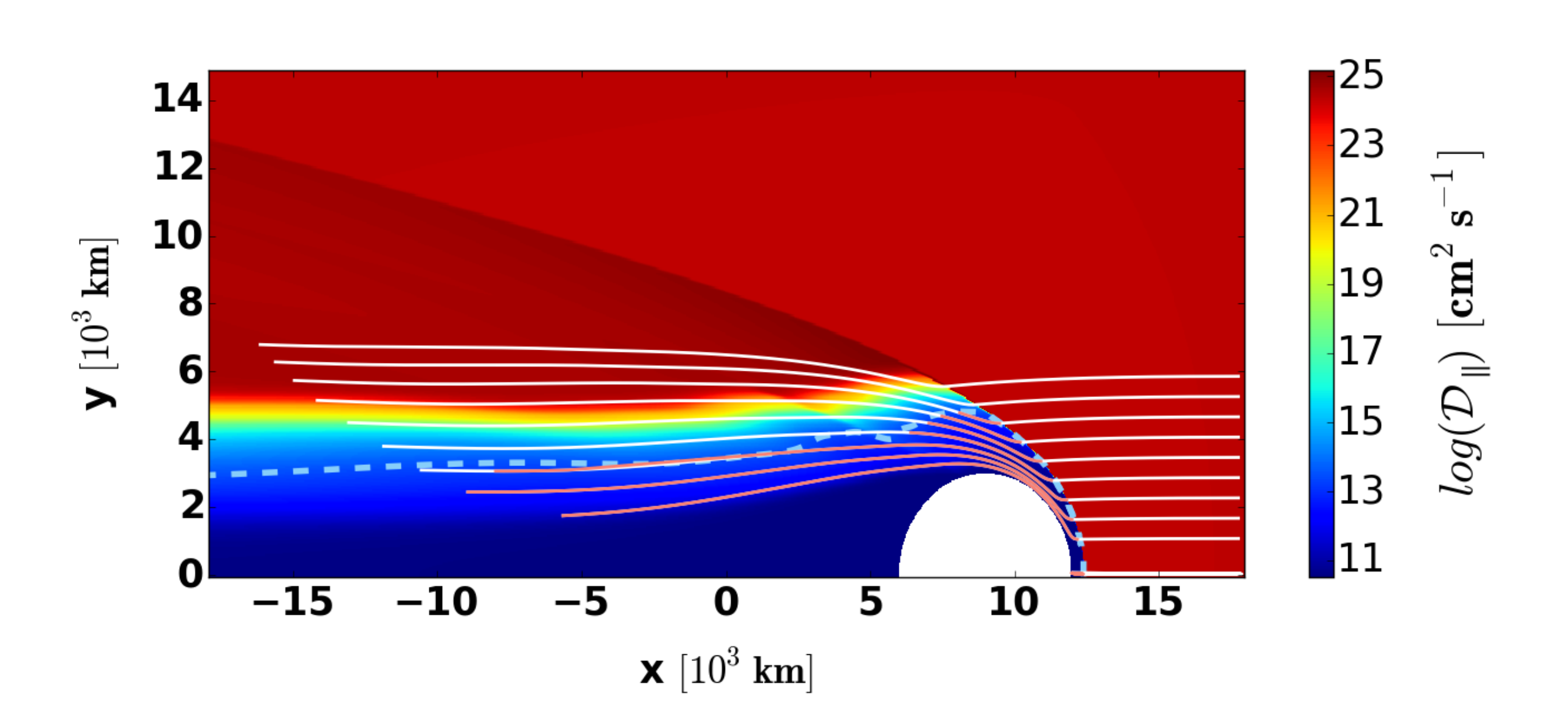}
\includegraphics[scale=0.55]{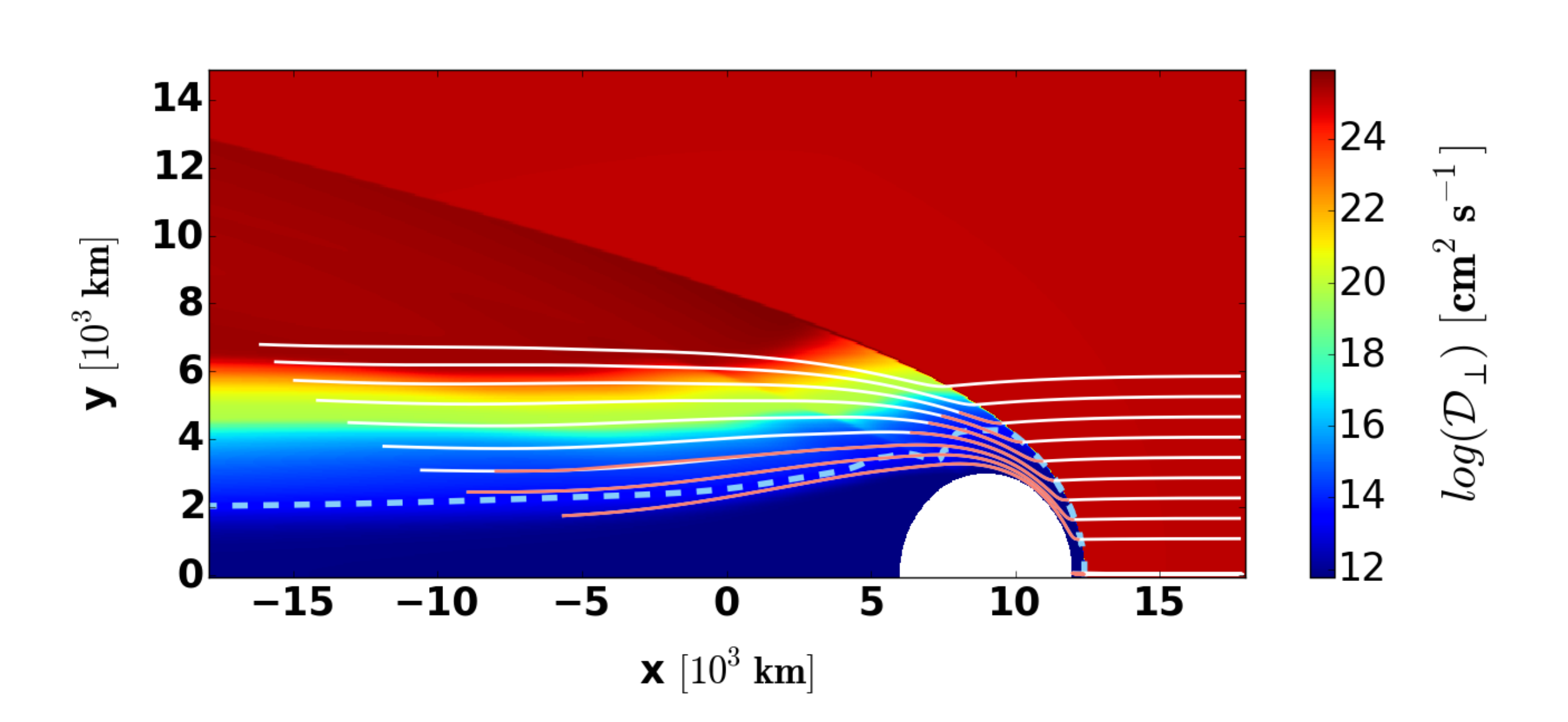}
\caption{The magnetic diffusivity ${\cal D}_{\parallel}$ (top) and ${\cal D}_{\perp}$ (bottom) around the planetary embryo 
(drawn as white).  The blue dashed line depicts the critical diffusivity threshold  
${\cal D} = 3 \times 10^{13} \, {\rm cm}^2\ {\rm s}^{-1}$. Above the dashed line the magnetic field is considered highly diffusive. The white solid lines are the simulated trajectories of chondrules under adiabatic conditions, with the parts where chondrule precursors are hotter than 1038 K (the kamacite Curie point) colored as light orange. On the whole, the majority of chondrules cool down from the Curie point in regions beyond the critical threshold and record the background magnetic field. See discussions in the text.
\label{fig:diffu}}
\end{figure*}

For the parallel conductivity, it is straightforward to show that the conductivity from electrons dominates, 
in which case the diffusion coefficient reduces to a much simpler form: 
\begin{equation} 
{\cal D}_{\parallel} = {\cal D}_{\rm OD}
 = \frac{ 9060 }{ x_{\rm e} } \, \left( \frac{ T }{ 1500 \, {\rm K} } \right)^{1/2} \, {\rm cm}^{2} \, {\rm s}^{-1},
\label{eq:bb94}
\end{equation}
where $x_{\rm e} = n_{\rm e} / n_{\rm H2}$ \citep{Blaes1994}.
Unlike ${\cal D}_{\parallel}$, ${\cal D}_{\perp} = {\cal D}_{\rm OD} + {\cal D}_{\rm AD}$ depends on the magnetic field strength.
In fact, to first order, ${\cal D}_{\rm AD} \propto B^2$. 
In the calculations that follow, we assume $B = 0.5 \, {\rm G}$, consistent with the paleofield measured by 
\cite{Fu2014}; larger magnetic fields would lead to greater rates of magnetic diffusion. 
We note that 0.5 G is a relatively weak magnetic field, resulting in a plasma beta 
$\beta = 8\pi P / B^2 \sim 10^3$ for $P \sim 10^{-5} \, {\rm bar}$.
The magnetic diffusion coefficients are depicted in Fig.~\ref{fig:diffu}, which shows ${\cal D}_{\parallel}$ and
${\cal D}_{\perp}$ behind the bow shock.
Also depicted is the rough boundary where ${\cal D} < 3 \times 10^{13} \, {\rm cm}^{2} \, {\rm s}^{-1}$. 

\begin{figure*}[ht!]
\centering
\includegraphics[scale=0.55]{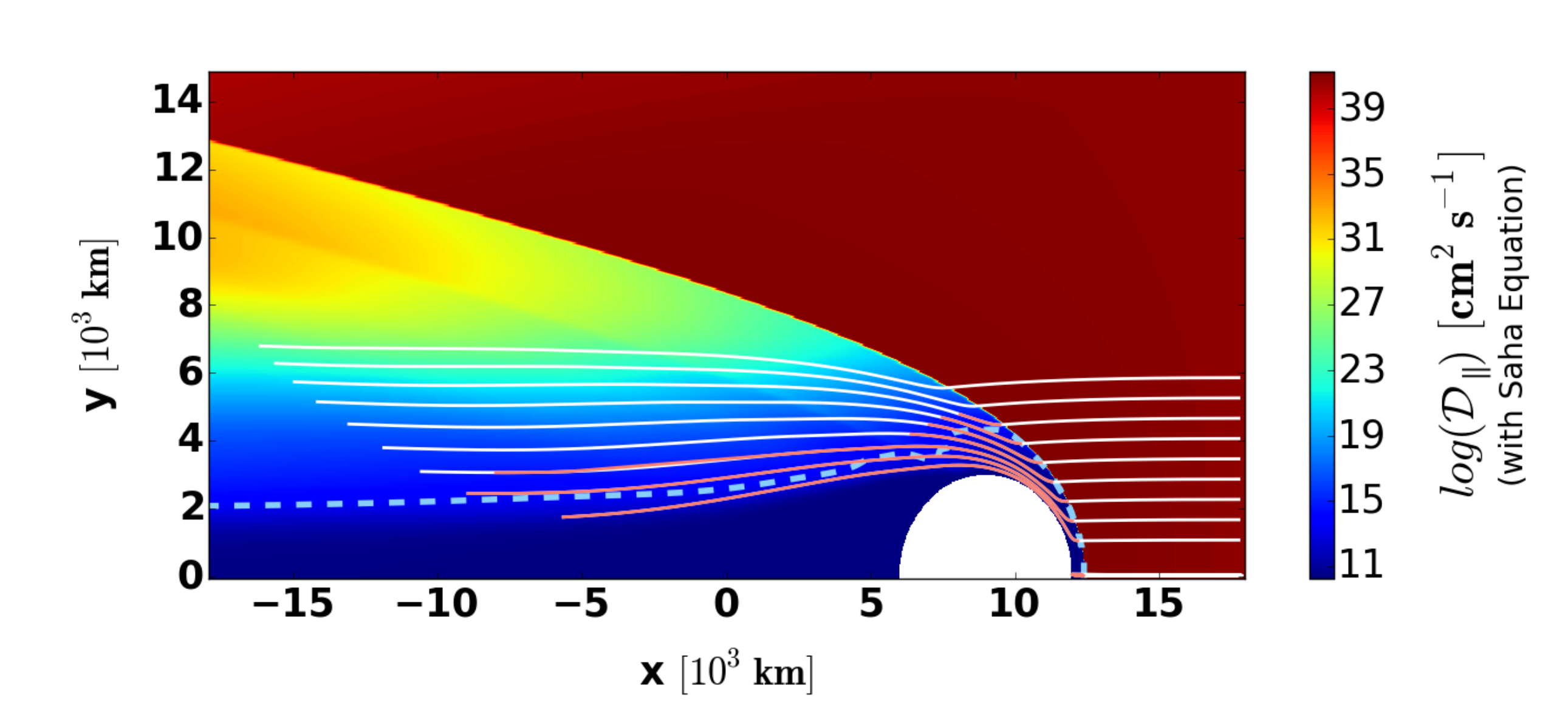}
\caption{The magnetic diffusivity ${\cal D}_{\parallel}$ around the planetary embryo calculated using the Saha equation (ignoring dust effects on ionization). Lines and symbols have the same meanings as in Fig.~\ref{fig:diffu}. See text for discussions.
\label{fig:diffu_saha}}
\end{figure*}

\subsection{Results}
\label{subsec:results}

The magnetic diffusion coefficients provide a rough idea of whether the magnetic field will remain amplified past the bow shock, 
or whether it will relax to the background field.
The magnetic diffusion timescale, the time it takes for magnetic flux to diffuse laterally into the unshocked region, is 
\begin{equation}
t_{\rm diff} \sim \frac{ L^2 }{ {\cal D} },
\end{equation}
where ${\cal D}$ pertains to either ${\cal D}_{\parallel}$ or ${\cal D}_{\perp}$, and $L \sim R_{\rm p} \sim 3000 \, {\rm km}$ 
is a relevant lengthscale. 
For comparison, the dynamical timescale for gas to flow past the planetary body is 
\begin{equation}
t_{\rm dyn} \sim \frac{ L }{ V },
\end{equation}
where $V$ is the post-shock velocity of gas and chondrules. 
If $t_{\rm diff} < t_{\rm dyn}$, then the magnetic field has time to relax to the background state by the time the 
chondrules reach the Curie point. 
That is, if ${\cal D}$ exceeds a critical value ${\cal D}_{\rm crit} \sim L V$, then chondrules will record a magnetic field relaxed 
to the background state.  
Using $L \sim 3000 \, {\rm km}$ and $V \sim 1 \, {\rm km} \, {\rm s}^{-1}$ yields 
${\cal D}_{\rm crit} \sim 3 \times 10^{13} \, {\rm cm}^{2} \, {\rm s}^{-1}$. 
To be more precise, the dynamical timescale is the time it takes chondrules to reach their Curie point, or drop in temperature
by about 1000 K. The cooling rates of chondrules in planetary bow shocks are $\approx 1000 - 2000 \, {\rm K} \, {\rm hr}^{-1}$,
meaning that chondrules will reach their Curie points after traveling downstream for about 
$t_{\rm dyn} = 30-60$ minutes, for a distance of about 2400-4800 km. 
On the other hand, the distance magnetic flux must travel laterally is, as is apparent from Fig.~\ref{fig:diffu},
less than 3000 km. 
On the whole, we judge ${\cal D}_{\rm crit} \sim 3 \times 10^{13} \, {\rm cm}^{2} \, {\rm s}^{-1}$ to be an accurate 
threshold. 

In Fig.~\ref{fig:diffu}, the blue dashed lines mark the threshold ${\cal D} = {\cal D}_{\rm crit}$ for both ${\cal D}_{\parallel}$ and ${\cal D}_{\perp}$. 
Therefore, in regions above the dashed lines, magnetic flux has sufficient time to diffuse laterally and relax to the background magnetic field strength. 
The fate of the magnetic field in the hot regions is less clear: it may or may not be amplified behind the shock, depending on the geometry. 
A behavior sometimes seen in the simulations of \cite{Boley2013} is that shocked gas sometimes cools and then heats up again because it enters the hot region behind the planetary embryo.
If the gas cooled sufficiently for tens of minutes, magnetic flux may diffuse out of that gas, relaxing to the background field, before it becomes frozen in the gas again.
A calculation of the magnetic field behind the planetary embryo is beyond the scope of this paper.
What can be said with certainty is that in all other regions, where ${\cal D} > {\cal D}_{\rm crit}$, the magnetic field
almost certainly relaxed to the background field of the nebula. 

All that remains is to show what region chondrules are in when they reach their Curie points and begin to record the local magnetic field.
In Fig.~\ref{fig:diffu}, we overlay the simulated trajectories of chondrules with the magnetic diffusivity maps.
The trajectories are colored light orange where the chondrules are heated up to above 1038 K, the Curie point of kamacite. 
The points where the orange trajectories become white again in the post-shock region mark where chondrules cool down to their Curie points.
It is quickly seen that under the action of Ohmic dissipation alone, most chondrules cool down at the edge of the low diffusivity region.
Although it may seem tricky to judge whether the chondrules record an enhanced magnetic field or the background field, or a field strength in between at a first glance,
we conclude that it is safe to state that most chondrules are magnetized beyond the critical diffusivity threshold line.
For one, chondrules record the field during cooling from the Curie point continuously down to ambient space temperatures. 
Even though the locations of the kamacite Curie points overlap with where the critical diffusivity threshold line lies,
the majority of chondrules would largely record a diffused magnetic field after passing this line, as they cool from the Curie temperature down to the ambient temperature.
For another, due to the existence of the background magnetic field, ambipolar diffusion also plays a non-negligible role in diffusing the magnetic field.
Under the combined action of Ohmic dissipation and ambipolar diffusion, it is clear that the vast majority of chondrules reach 1038 K beyond the critical diffusivity threshold in the highly diffusive region.
Even at the relatively weak magnetic field strength of 0.5 G, ambipolar diffusion may significantly enhance the rate of magnetic 
flux diffusion.
Note that in these simulations we have assumed the chondrules are in thermal equilibrium with the surrounding gas. 
In reality, chondrules might experience effective radiative cooling and cool down at higher rates in the post-shock region. 
But even if the temperature difference between gas and chondrules is as large as 100 K, the results would not be changed much and the above discussions remain valid.
Moreover, 1038 K is the highest Curie temperature of minerals found in chondrules, yet still not high enough for the gas to maintain an ionization fraction large enough to "freeze" the magnetic field.
Lower Curie points, like those of taenite ($<$ 873 K), martensite magnetite (853 K), and pyrrhotite (600 K), would be reached further beyond the critical diffusivity threshold line.

Fig.~\ref{fig:diffu_saha} is essentially the same as Fig.~\ref{fig:diffu} top panel, except that the magnetic diffusivity $\cal D_{\parallel}$ is derived from ionization fractions calculated using the Saha equation (Fig.~\ref{fig:ioniz} lower panel). Due to the lack of thermionic emission from dust grains, the ionization levels in regions with temperatures around 1000 K are much lower. Therefore, the flux-freezing region shrinks and the critical diffusivity threshold line moves downward, compared to Fig.~\ref{fig:diffu} (top panel). Chondrules are supposed most likely to record a diffused background field in such a scenario.

To test the sensitivity of our results, we also performed bow shock simulations under different sets of initial parameters. 
By increasing the relative velocity of the planetary embryo to disk gas to 10 km s$^{-1}$, 
we produce a post-shock region with larger high-temperature area. 
Both the critical diffusivity threshold lines and the locations of the Curie points of chondrules are correspondingly shifted further away from the planetary embryo. 
The conclusion that most chondrules would record a nebular magnetic field remains unchanged. 
We also altered the size of injecting chondrule precursors by one standard deviation from the original input ($\sim$ 0.03 cm), 
assuming they follow a log-normal size distribution \citep{Nelson2002}. 
The different sizes of chondrules change the survival rates of the population, but not the bulk pattern of their trajectories, 
and therefore not the locations of their Curie points. Our previous conclusion still holds.

On the whole, it is safe to say that chondrules spend almost all of their time, and certainly the time after they reach their Curie temperatures,
outside of the region where magnetic fields can be amplified. 
Chondrules melted in bow shocks should record the background field of the nebula.

We end this section by considering the implications of these calculations for the possibility 
that the chondrules might have recorded a magnetic field generated by the embryo. 
In particular, Recent paleomagnetic studies of a achondrites have found that early planetary bodies likely generated dynamo magnetic fields 
with surface paleointensities of $\sim$ 0.1-1 G \citep{Wang2017, Fu2012, Bryson2015}. 
Figs. \ref{fig:diffu} and \ref{fig:diffu_saha} show that the chondrules reach the kamacite Curie temperatures at a distance of $\sim$ 1.5 to $>$ 7 planetary radii from the embryo's center. 
If we assume that its dynamo field is dipolar, then scaling the paleointensities above shows that this would have 
occurred where the embryo's dynamo field was somewhere between $\sim$ 10$^{-5}$ to 0.3 G. 
Even the latter maximum value is about a factor of $\sim$ 2 lower than that recorded by the Semarkona chondrules. 
Furthermore, a comparison of Fig. \ref{fig:temp} with Figs. \ref{fig:diffu} and \ref{fig:diffu_saha} shows that the chondrules reach their 750 K blocking temperatures at a distance of $\sim$ 3 planetary radii, 
at which point the dynamo field would be $<$ 0.04 G, well below the 0.54 G paleointensity recorded by Semarkona.
We conclude that total thermoremanent magnetization in chondrules formed from cooling from the Curie point to ambient space temperatures is unlikely to be the product of a planetesimal dynamo for paleointensities exceeding a few tens of mG.
\\
\begin{figure*}[ht!]
\centering
\includegraphics[scale=0.7]{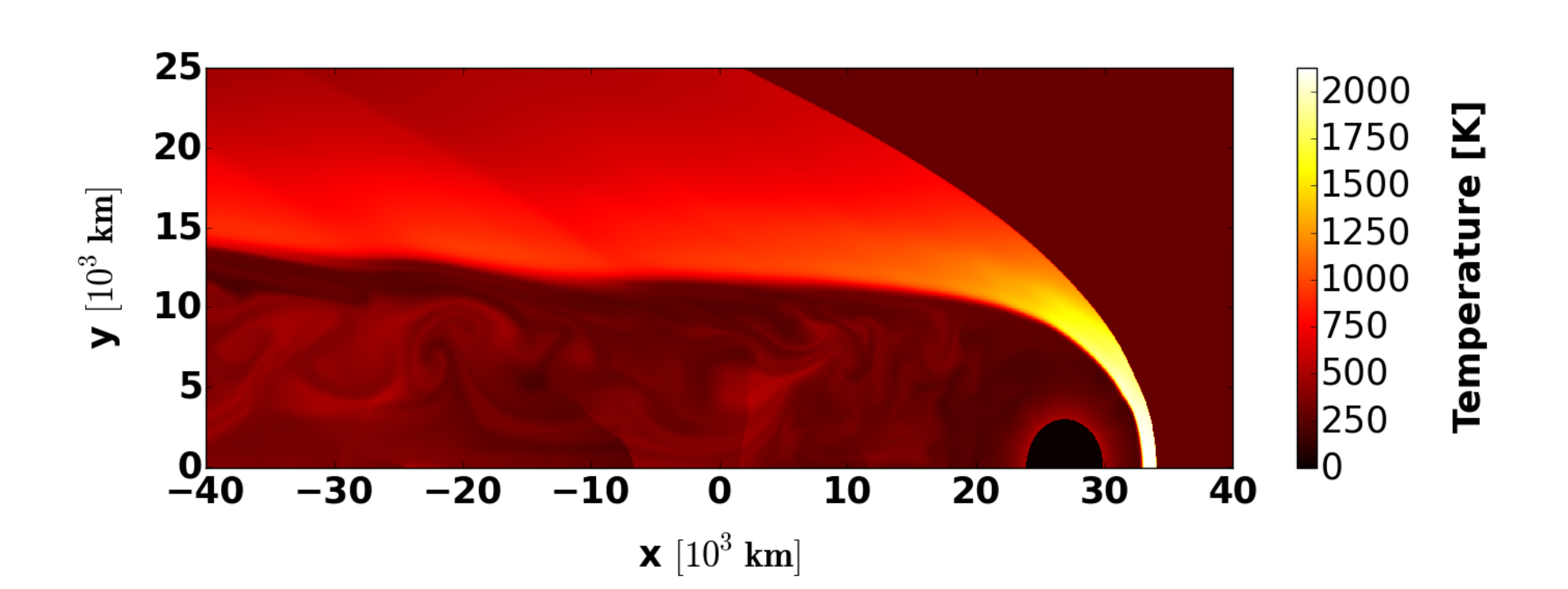}
\includegraphics[scale=0.7]{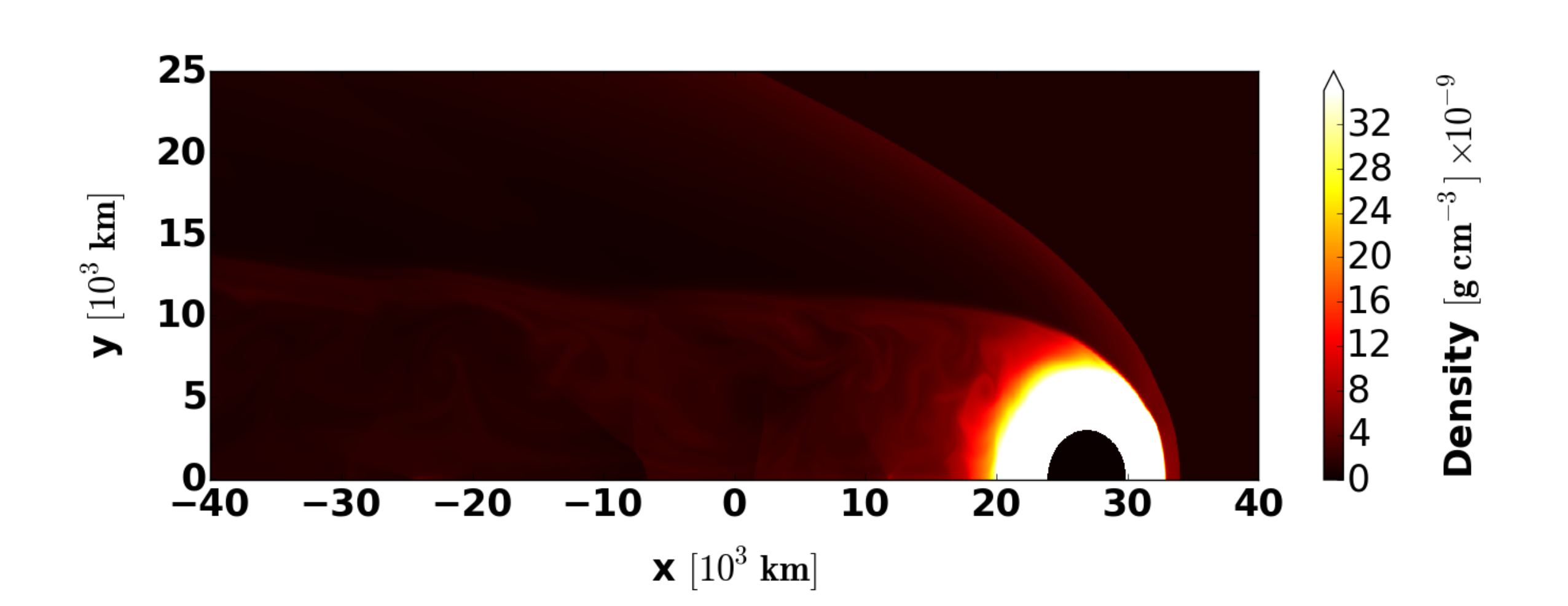}
\caption{Snapshots of representative gas temperature (top) and density (bottom) in the bow shock region where a thick planetary proto-atmosphere is present. The planetary embryo is drawn as black.  
\label{fig:txydxy_atm}}
\end{figure*}


\begin{figure*}[ht!]
\centering
\includegraphics[scale=0.7]{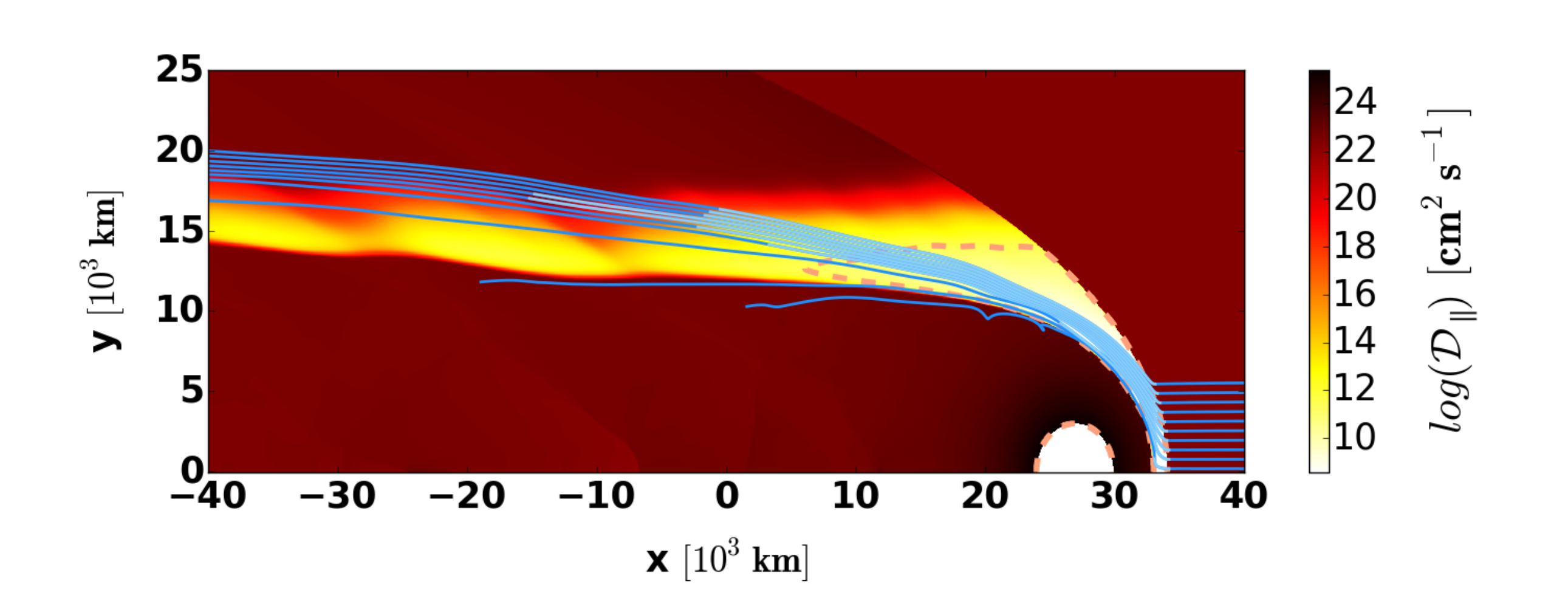}
\includegraphics[scale=0.7]{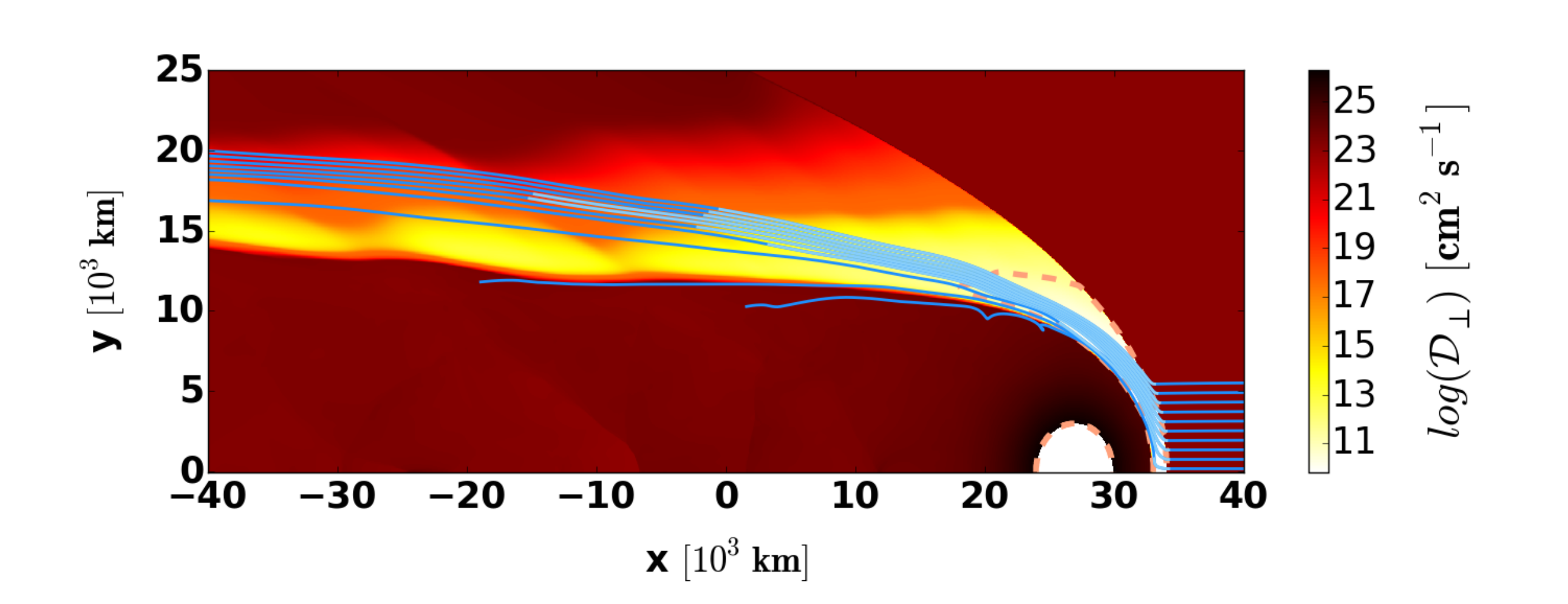}
\caption{The magnetic diffusivity ${\cal D}_{\parallel}$ (top) and ${\cal D}_{\perp}$ (bottom) around the planetary embryo (drawn as white) with a thick atmosphere. Note that the color scheme is different from Fig. \ref{fig:diffu} and adjusted for better visualization in the high-mass atmosphere case. The orange dashed line depicts the critical diffusivity threshold ${\cal D}=3\times 10^{13}\,  {\rm cm}^2\ {\rm s}^{-1}$. The amplified magnetic field in any regions above the dashed line is quickly diffused to the background level. The blue solid lines are the simulated chondrule trajectories, with the parts where chondrule precursors are hotter than 1038 K colored as light blue. The vast majority of chondrules cool down far beyond the critical threshold and record the background magnetic field.
\label{fig:diffu_atm}}
\end{figure*}

\subsection{The Influence of Planetary Proto-atmospheres}
\label{subsec:atmos}
As the planetary embryos are embedded in a gaseous protoplanetary disk, it is reasonable to believe that they are able to maintain proto-atmospheres either through the accretion of nebula gas or outgassing from the planetary surface, or both. 
Particularly, \cite{Stokl2015} has pointed out that Earth-size protoplanets can hold up to thousands of bars of hydrogen-rich atmospheres when nebula gas is dissipated. 
Although some atmosphere loss mechanisms have been proposed for planets in their early life 
(e.g. ram-pressure stripping; ultraviolet and soft X-ray (XUV) radiation-driven escape; hydrodynamical effect from disk evaporation), 
the timescales of these escape processes are probably not small enough, compared to the ones of atmosphere accretion and chondrule formation, to immediately remove the proto-atmospheres. 
We believe the planetary embryo could possess an atmosphere in at least some portion of its time with a bow shock.
We will revisit the topic later in the section.

The presence of an atmosphere could increase the shock size, modify the state variables, deflect chondrule trajectories and thus probably provide a different scenario of magnetic diffusion and chondrule cooling in the bow shock.
The amount of the atmosphere that the embryo retains is  subject to planet mass, the accretion and stripping rates and the windspeed conditions at different orbital phases, and therefore should vary over time (\citealt{Mann2016}). 
Here we explore the two extreme cases, the atmosphere-free case discussed above in \S~\ref{subsec:3dmodel} to \S~\ref{subsec:results} and the high-mass atmosphere case that will be discussed below, 
which should bracket any other possible cases where an atmosphere is present. 

In the high-mass atmosphere case, we place an atmosphere mostly composed of H$_2$ and He with polytropic structure, following a spherical symmetric density profile (\citealt{Morris2012}; \citealt{Mann2016}):
\begin{equation}
\rho(r)=\rho_0\Big[\, 1+H\Big(\frac{1}{r}-\frac{1}{R_e}\Big)\Big]^{1/(\gamma-1)}
\end{equation}
where $\rho_0$ is the surface gas density, 
$r$ is the distance to the center of the planet, 
$\gamma \sim 1.35$ is the adiabatic index, 
and $R_e$ is the planet radius. 
$H=\frac{GM(\gamma-1)}{\gamma K\rho_0^{\gamma-1}}$ functions similar to an atmosphere scale height, 
$K$ is the polytropic constant. 
The atmosphere has a similar set up as in \cite{Mann2016}, 
with a mass of about $2.7\times 10^{20}$ g, 
or 11 Martian atmospheres, and creates 14 mbar of surface pressure. 
The simulations are extended to include 800 cells along $x$ axis and 3200 cells along $y$ axis to better capture the physics.
Fig. \ref{fig:txydxy_atm} shows the representative temperature and gas density around the embryo with a thick atmosphere, as a comparison to Fig. \ref{fig:temp}.

The existence of the atmosphere generates a cooler and denser boundary layer between the protoplanet and the post-shock gas flow.
At this boundary, which reaches 13,000 km in radius behind the planetary embryo, vigorous turbulence including Kelvin-Helmholtz (KH) instabilities develop.
As pointed out in \cite{Morris2012} and \cite{Ouellette2007}, KH instabilities can strip away the atmosphere with an efficiency $\sim 1\%$ 
(relative to the incident mass flux), translating to an atmospheric mass loss rate $\sim 10^{12}\ {\rm g}\ {\rm s}^{-1}$.
In their numerical simulations, \cite{Mann2016} measured a stripping rate of $\sim 5\times 10^{14}\ {\rm g} \ {\rm s}^{-1}$, which
is considerably higher.  
Both rates are lower than the estimated Bondi accretion rate of a stationary atmosphere by one to three orders of magnitude. 
However, in reality, atmosphere accretion rate is also affected by the relative motion between the protoplanet 
and the nebula gas and probably the bow shock structure itself. 
There are different phases of mass accretion and mass loss of the atmosphere depending on the details of the relative wind speeds as a function of embryo orbit.
The impact of an eccentric orbit on atmosphere accretion is beyond the scope of this study,
but remains as a goal of our future work.

Following the same methodology layed out in \S~\ref{subsec:ionizn} and \S~\ref{subsec:diffu},
we calculate the magnetic diffusivity ${\cal D}_{\parallel}$ and ${\cal D}_{\perp}$ for the high-mass atmosphere case (Fig. \ref{fig:diffu_atm}). 
Unlike the atmosphere-free case,
the thick atmosphere creates a large low-temperature high-diffusivity cylinder with a radius of 1,3000 km behind the planet.
The atmosphere also deflects the chondrule trajectories so drastically that 
very few chondrules that enter with low impact parameter make it into the high-diffusivity cylinder.
Revolving trajectories are observed for these few chondrules under the influence of KH instability.
It is clear that under either Ohmic dissipation alone or both Ohmic and ambipolar diffusion, 
the vast majority of chondrules cool down to their Curie points in the downstream of the post-shock region 
far away from the critical threshold contour line and are magnetized by the background nebula field.
Again, considering the higher cooling rates of chondrules in reality, a temperature gap as large as 100 K between the gas and chondrules only alters the above results subtly.
We conclude that the presence of an atmosphere significantly increases the chance that chondrules cool down in highly diffusive post-shock regions,
and therefore  and record the background magnetic field.

We can also consider the possibility that chondrules were magnetized by a dynamo generated on the atmosphere-bearing embryo.  Fig. \ref{fig:diffu_atm} shows that in this case, the chondrules reach the kamacite Curie point at $\sim$ 15 planetary radii from the embryo.  Again assuming a dipolar dynamo with surface intensity of 0.1 to 1 G, we see that the dynamo field at this distance is negligible ($<$ 3$\times$10$^{-4}$ G) and so cannot explain the Semarkona chondrule paleointensities.
\\
\section{Discussion and Conclusions} 
\label{sec:condis}

In this article we have determined the magnetic fields that would be recorded by chondrules melted by nebular shocks.
In the case of large-scale shocks, such as those generated by gravitational instabilities in the protoplanetary disk,
we find that magnetic fields parallel to the shock front are amplified by factors $\chi \sim 10-30$ regardless of the 
rate of magnetic diffusion, although magnetic fields normal to the shock front are not amplified. 
We also have considered the rate of magnetic diffusion in the gas behind a planetary embryo bow shock. 
We have used the output from the radiation hydrodynamics simulations to constrain the gas densities,
and temperatures, and the trajectories of chondrules, in the post-shock region. 
We have calculated the ionization state of the post-shock gas using the formalism of \cite{Desch2015},
and computed the coefficients of magnetic diffusion due to Ohmic dissipation and ambipolar diffusion.
In the case of an atmosphereless planetary embryo, we find that these coefficients typically exceed ${\cal D}_{\rm crit} \approx 3 \times 10^{13} \, {\rm cm}^{2} \, {\rm s}^{-1}$
and will relax to the background field, everywhere except the region near and behind the planetary body.
If we consider only the effects of Ohmic dissipation, most chondrules have already cooled to their Curie points upon arriving at the critical diffusivity threshold.
With the effects of ambipolar diffusion included, the majority cool down in regions where ${\cal D} > {\cal D}_{\rm crit}$.
Therefore, during their subsequent cooling to ambient temperatures, which is the the time when they acquire remanent magnetization, they record the ambient (i.e., non-amplified) nebular field.
On the whole, most chondrules are magnetized in the background magnetic field in the bow shocks produced by atmosphere-free protoplanets.
In another extreme case where the planetary body possesses a thick atmosphere ($2.7\times 10^{20}\ {\rm g}$), ${\cal D} < {\cal D}_{\rm crit}$ except 
in a 13,000-km radius cylinder behind the planet 
and in the post-shock region far from the planet.
Almost all the trajectories of chondrules lie outside of the highly diffusive cylinder. 
Yet the time and place they cool to their Curie points are far away enough in the downstream so they end up recording the background magnetic field anyway.
As such, chondrules melted in planetary embryo bow shocks record magnetic fields in regions that have relaxed to the background 
magnetic field.
We also show that the chondrules cool to ambient space temperatures sufficiently far from the embryos that they are unlikely to be substantially magnetized by possible dynamo fields generated in the embryos' metallic cores.

We can approximate the temperature below which ${\cal D} > {\cal D}_{\rm crit}$ and the rate of magnetic diffusion is rapid enough
that the magnetic field relaxes to the background level.
The temperature at the Curie point is 1038 K, and this is the relevant temperature at which to calculate the rate of magnetic diffusion.
The ionization of the gas is dominated by thermionic emission, not collisional (thermal) ionizations, but we can approximate the 
behavior by using the Saha equation and assuming only Ohmic dissipation.
In that case, the number density of ${\rm K}^{+}$ ions is given by the Saha equation and, assuming $n_{\rm e} = n_{{\rm K}^{+}}$, 
the number density of electrons is 
\begin{equation}
\begin{split}
x_{\rm e} = & \frac{ n_{\rm e} }{ n_{\rm H2} } = \left( \frac{ n_{\rm K0} / n_{\rm H2} }{ n_{\rm H2} } \right)^{1/2} \, \cdot
\left( \frac{ 2 g_{+} }{ g_0 } \right)^{1/2} \, \\
\cdot & \left( \frac{ 2\pi \, m_{\rm e} \, k T }{ h^2 } \right)^{3/4} \, \cdot \exp \left( -\frac{\rm IP}{ 2 k T } \right) \\
= & 0.65 \, \cdot \left( \frac{ n_{\rm H2} }{ 10^{14} \, {\rm cm}^{-3} } \right)^{-1/2} \, \cdot
 \left( \frac{ T }{ 1500 \, {\rm K} } \right)^{3/4} \, \\
\cdot & \exp \left( -\frac{ 16.782 }{ T / 1500 \, {\rm K} } \right)
\end{split}
\end{equation}
Combined with Eq.~\ref{eq:bb94}, and assuming $n_{\rm K0} / n_{\rm H2} \approx 3 \times 10^{-7}$ (Lodders 2003), it is 
quickly seen that ${\cal D}_{\rm OD} = {\cal D}_{\rm crit}$ at a temperature of about 1175 K.
At $T \approx 1100 \, {\rm K}$, using the Saha equation, ${\cal D}_{\rm OD} \approx 1.3 \times 10^{14} \, {\rm cm}^{2} \, {\rm s}^{-1}$, 
and the timescale to magnetic flux to diffuse laterally about 2000 km is about 5 minutes, about the same length of time for chondrules
to cool from 1175 K to 1038 K (at a cooling rate of $2000 \, {\rm K} \, {\rm hr}^{-1}$). 
By considering the ionization fraction due to thermionic emission, and including the effects of ambipolar diffusion, very similar 
results are obtained.

\cite{Fu2014} inferred paleofields of 0.54 G from chondrules in the L3.0 Semarkona chondrite. 
If these chondrules were melted by large-scale solar nebula shocks, then they would have recorded magnetic fields an
order of magnitude greater than the  background magnetic field, implying background fields $\sim 0.02 - 0.05 \, {\rm G}$.
However, gravitational instabilities that might drive large-scale shocks require a high surface density that is most likely early
in the evolution of the protoplanetary disk, whereas chondrules in ordinary chondrites apparently formed at a protoplanetary disk age 
$\approx 1.5 - 3$ Myr \citep{Villeneuve2009}.
We view it as more likely that the chondrules in Semarkona were melted by passage through bow shocks in advance of planetary 
embryos on eccentric orbits (potentially due to resonant interactions with a recently formed massive Jupiter).
In that case, the background magnetic field of the solar nebula is likely to be close to the paleofield recorded by the 
chondrules, $\approx 0.5$ G. 

A background magnetic field of the solar nebula disk $\approx 0.5$ G compares well with theoretical estimates of the maximum
fossil magnetic field in the region where chondrites formed ($\approx 2-3$ AU). 
\cite{Nakano1986a, Nakano1986b} estimated field strengths $\sim 1$ G in a region of size $\sim 4$ AU. 
\cite{Desch2001} calculated field strengths $\sim 0.1$ G over a region of size $\sim 20$ AU during prestellar collapse,
the exact value depending on parameters such as grain size and cosmic-ray ionization rate.
If grains coagulated to radii $> 1 \, \mu{\rm m}$ during collapse, the magnetic field in this region would be $> 0.5$ G
(see also \citealt{Zhao2016}). 
Grain growth to this size during the molecular cloud stage is predicted \citep{Mouschovias1999}. 
\cite{Masson2016} likewise found that magnetic fields are not amplified above about 0.1 G within tens of AU. 
\cite{Kunz2010} performed similar calculations and found field strengths $\sim 0.2$ G over regions $\sim 2$ AU in radius.
These calculations all predict magnetic fields during prestellar collapse of a few tenths of a Gauss within the region where 
chondrules and chondrites will subsequently form.

A background magnetic field of the solar nebula disk $\approx 0.5 \, {\rm G}$ is also consistent with the magnetic field
needed to transport angular momentum in protoplanetary disks.
If angular momentum and mass are radially transported by the MRI, then the magnetic field strength
relates to the mass accretion rate as 
\begin{equation}
B > 0.3 \, \left( \frac{ \dot{M} }{ 10^{-8} \, M_{\odot} \, {\rm yr}^{-1} } \right)^{1/2} \, 
 \left( \frac{ r }{ 2.5 \, {\rm AU} } \right)^{-11/8} \, {\rm G} 
\end{equation}
\citep{Fu2014}.
Conversely, if angular momentum is transported by a magnetocentrifugal disk wind, then the magnetic field strength is 
related to mass accretion rate as 
\begin{equation}
B > 0.03 \, \left( \frac{ \dot{M} }{ 10^{-8} \, M_{\odot} \, {\rm yr}^{-1} } \right)^{1/2} \, 
 \left( \frac{ r }{ 2.5 \, {\rm AU} } \right)^{-5/4} \, {\rm G} 
\end{equation}
\citep{Fu2014}.
We favor minimal amplification of the magnetic field during chondrule formation, meaning that the magnetic fields recorded
by Semarkona chondrules were sufficient to transport mass by either the MRI or by disk winds.

\acknowledgements
Acknowledgements -   
We would like to thank the anonymous referee for his/her careful reading and constructive comments and suggestions that helped us improve this paper greatly.
We are grateful to the support from NASA Emerging Worlds (PI Benjamin Weiss, NNX15AH72G).
ACB's contribution was supported by an NSERC Discovery grant, The University of British Columbia, the Canadian Foundation for Innovation, and the BC Knowledge Development Fund.
\\

\end{document}